\documentclass{ifacconf}

\usepackage{multirow}
\usepackage{graphicx}      
\usepackage{amsmath}
\usepackage{bm}
\usepackage{natbib}        
\usepackage{filecontents}

\makeatletter
\let\old@ssect\@ssect 
\makeatother

\usepackage{natbib}
\usepackage{hyperref}

\makeatletter
\def\@ssect#1#2#3#4#5#6{%
  \NR@gettitle{#6}
  \old@ssect{#1}{#2}{#3}{#4}{#5}{#6}
}
\makeatother
\begin{document}
\begin{frontmatter}

\title{Satellite Dynamics Toolbox Library: a tool to model multi-body space systems for robust control synthesis and analysis} 

\author[First]{Francesco Sanfedino} 
\author[First]{Daniel Alazard} 
\author[Second]{Ervan Kassarian}
\author[Third]{Franca Somers}

\address[First]{ISAE-SUPAERO, Université de Toulouse, Toulouse, France}
\address[Second]{DyCSyT S.A.S, Toulouse, France}
\address[Third]{ONERA, The French Aerospace Lab, Toulouse, France\\(emails:[francesco.sanfedino/daniel.alazard]@isae.fr, ervan.kassarian@dycsyt.com, franca.somers@onera.fr)}

\begin{abstract}                
The level of maturity reached by robust control theory techniques nowadays contributes to a considerable minimization of the development time of an end-to-end control design of a spacecraft system. The advantage offered by this framework is twofold: all system uncertainties can be included from the very beginning of the design process; the validation and verification (V\&V) process is improved by fast detection of worst-case configurations that could escape to a classical sample-based Monte Carlo simulation campaign. 
Before proceeding to the control synthesis and analysis, a proper uncertain plant model has to be available in order to push these techniques to their limits of performance. In this spirit, the Satellite Dynamics Toolbox Library (SDTlib) offers many features to model a spacecraft system in a multi-body fashion on \texttt{SIMULINK}. Parametric models can be easily built in a Linear Fractional Transformation (LFT) form by including uncertainties and varying parameters with minimal number of repetitions. Uncertain Linear Time Invariant (LTI) and uncertain Linear Parameter-Varying (LPV) controllers can then be synthesized and analyzed in a straightforward way. The authors present in this article a tutorial, that can be downloaded at \url{https://nextcloud.isae.fr/index.php/s/XDfRfHntejHTmmp}, to show how to deal with an end-to-end robust design of a spacecraft mission and to provide to researchers a benchmark to test their own algorithms.
\end{abstract}


\end{frontmatter}

\section{Introduction}

End-to-end spacecraft control design in industry is a long process that starts with the modeling of the spacecraft dynamics and ends up with the V\&V campaigns. This process is not sequential and several iterations are necessary before system certification occurs. The major difficulty consists in assuring that the system will behave correctly without loss of stability and performance in presence of many system uncertainties. The LFT framework, developed from the early eighties (\cite{doyle1982analysis}), offers the possibility to design robust control laws in spite of parametric and dynamic uncertainties in the system and to analyze then the closed-loop stability and worst-case (WC) performance.
In order to put in place these techniques, now available in the Robust Control Toolbox of \texttt{MATLAB} (\cite{RCT_user_guide}), an LFT model of the plant has to be built. The number of uncertain/varying parameters' repetitions affects the efficiency of the synthesis/analysis algorithms and has to be kept minimal. The Satellite Dynamics Toolbox Library (SDTlib) (\cite{sdt}), based on the Two-Input Two-Output Port (TITOP) theory (\cite{titop}), allows the user to build a multi-body flexible spacecraft in the LFT framework by assembling of elementary blocks in \texttt{SIMULINK}. Simple bodies (rigid bodies, flexible beams, flexible plates), complex Finite Element Model (FEM) structures (that can be imported from \texttt{NASTRAN} analysis files), mechanisms (reaction wheels, solar array drive mechanisms) and other flexible dynamics (i.e. sloshing effects) are available in this library.

The main objective of this paper is to present a demo of a complete design process of an attitude control system (ACS) for a spacecraft mission by combining the SDTlib together with the features available in the Robust Control Toolbox, the SMAC library (\cite{biannic2016smac}) and the STOWAT library (\cite{ThRoBi19_acc}). This benchmark is destined to researchers and industrial actors who would like to be introduced to the proposed techniques and/or to test their own robust control/analysis algorithms.
The results presented in this paper were run in MATLAB R2021b on a MacBook Pro 2 GHz Intel Core i5.
In the following parts of the article, Section \ref{sec:modeling} will detail the spacecraft dynamics modeling. A robust controller synthesis is then outlined in Section \ref{sec:control}. Section \ref{sec:analysis} presents the robust analysis of the system and finally some conclusions and future perspectives are provided in Section \ref{sec:conclusion}.

\section{Benchmark modeling}
\label{sec:modeling}

\begin{figure}[ht!]
	\centering
	\includegraphics[width=.7\columnwidth]{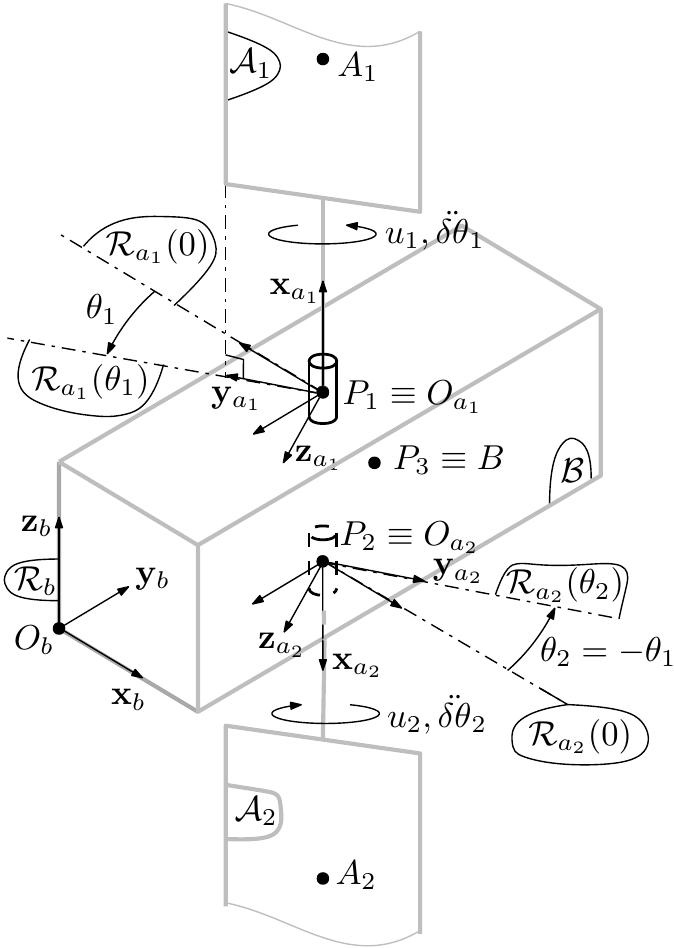}
	\caption{Spacecraft geometry.}
	\label{fig:spacecraft_geometry}
\end{figure}

The spacecraft $\mathcal{S}\mathcal{C}$ in Fig. \ref{fig:spacecraft_geometry} is composed of: the main body $\mathcal{B}$ with its center of mass (CoM) $B$, its reference point $O_b$ and its body frame ${\mathcal{R}}_b(O_b;\mathbf{x}_b,\mathbf{y}_b,\mathbf{z}_b)$; two symmetrical flexible solar arrays ${\mathcal{A}}_1$ and ${\mathcal{A}}_2$ cantilevered to $\mathcal{B}$ at the points $P_1$ and $P_2$ with an angular configuration $\theta_1$ and $\theta_2$, respectively. $A_i$, $O_{a_i }$ and ${\mathcal{R}}_{a_i }(O_{a_i};\mathbf{x}_{a_i},\mathbf{y}_{a_i},\mathbf{z}_{a_i})$ are respectively the CoM, the reference point and the body frame of ${\mathcal{A}}_i$ ($i=1,\,2$).
The angular configurations of the two solar panels are symmetrical: $\theta_1 =\theta$ and $\theta_2 =-\theta$. In the Figure, ${\mathcal{R}}_{a_i } (0)$ and ${\mathcal{R}}_{a_i } (\theta_i )$ represent two geometric configurations of ${\mathcal{R}}_{a_i }$ for the nominal configuration ($\theta_i =0$) and for a given angle $\theta_i$.
$\left[\mathbf{D}_B^{\mathcal{S}\mathcal{C}}\right]_{{\mathcal{R}}_b }^{-1} ({\mathrm{s}},\theta )$ is the model of the spacecraft $\mathcal{S}\mathcal{C}$ at the CoM $B$ of the main body $\mathcal{B}$ and projected in the main body frame axes ${\mathcal{R}}_b$ for a given angular configuration $\theta$, that is the $6\times 6$ transfer between:
\begin{itemize}
\item the resultant external wrench $\left[\mathbf{W}_{ext/\mathcal{B},B} \right]_{{\mathcal{R}}_b }$ (6 components: 3 forces and 3 torques) applied to $\mathcal{B}$ at the point $B$,
\item the dual vector of acceleration $\left[{{\ddot{\mathbf x} }}_B \right]_{{\mathcal{R}}_b }$ of point $B$ (6 components: 3 translations, 3 rotations).
\end{itemize}

\subsection{Main Body}
This body can be simply modeled with the block \texttt{\textbf{multi port rigid body}} of the sub-library \texttt{\textbf{6 dof bodies}} of the SDTlib. Please refer to Table \ref{tab:data} for all benchmark data. The model $\left[\mathbf D_{P_1 ,P_2 ,P_3 }^{\mathcal{B}}\right]^{-1}$ defines the transfer functions among the wrenches and the accelerations of the three connection points $P_1$, $P_2$ and $P_3$ on body $\mathcal{B}$.

\subsection{Direction Cosine Matrix}
The Direction Cosine Matrix (DCM):
\begin{equation}
	\mathbf P_{a_1 (0)/b} =\left[\begin{array}{ccc}
		0 & -1 & 0\\
		0 & 0 & -1\\
		1 & 0 & 0
	\end{array}\right]
\end{equation} 
defines the rotation:
\begin{itemize}
	\item from the body ${\mathcal{A}}_1$ frame: ${\mathcal{R}}_{a_1 } (0)$ in the nominal configuration ($\theta =0$),
	\item to the body $\mathcal{B}$ frame: ${\mathcal{R}}_b$. 
\end{itemize}

This matrix expresses the coordinates of $\mathbf x_{a_1} (0)$, $\mathbf y_{a_1} (0)$ and $\mathbf z_{a_1} (0)$ in ${\mathcal{R}}_b$.
In the same way: 
\begin{equation}
	\mathbf P_{a_2 (0)/b} =\left[\begin{array}{ccc}
		0 & 1 & 0\\
		0 & 0 & -1\\
		-1 & 0 & 0
	\end{array}\right]
\end{equation}
These DCMs, twiced $\mathbf P_{a_2 (0)/b}^{\times2} = \mathrm{blkdiag}\left(\mathbf P_{a_2 (0)/b},\mathbf P_{a_2 (0)/b}\right)$ (one for the translation, one for the rotation), can be modeled with the block \texttt{\textbf{6x6 DCM}} of the sub-library \texttt{\textbf{3/6 dof DCM}} of the SDTlib.

Note that the direct DCMs act on the inputs of the model $\left[\mathbf D_{P_1 ,P_2 ,P_3 }^{\mathcal{B}}\right]_{{\mathcal{R}}_b }^{-1}$ while the DCMs transposed are on the outputs of this model.

The DCM between frames ${\mathcal{R}}_{a_1 } (0)$ and ${\mathcal{R}}_{a_1 } (\theta )$ is associated to the rotation of a varying angle $\theta$ around the axis $\mathbf x_{a_1}$:
\begin{equation}
	\mathbf P_{a_1 (\theta )/a_1 (0)} = \mathbf R(\theta ,\mathbf x_{a_1 } )=\left\lbrack \begin{array}{ccc}
		1 & 0 & 0\\
		0 & \cos \theta  & -sin\theta \\
		0 & \sin \theta  & \cos \theta 
	\end{array}\right\rbrack
\end{equation}

In the same way:
\begin{equation}
	\mathbf P_{a_2 (\theta )/a_2 (0)} = \mathbf R(\theta ,-\mathbf x_{a_2 } )
\end{equation}
These DCMs can be modeled with the \texttt{\textbf{6x6 v-u\_rotation}} of the sub-library \texttt{\textbf{3/6 dof DCM}} of the SDTlib:
Note that the angle $\theta$ is declared as an uncertain parameter (\texttt{\textbf{ureal}}) varying between $-\pi$ and $\pi$.  This block uses the parameterisation $\sigma_4 =\tan (\theta /4)$ to express the DCM as a Linear Franctional Transformation (LFT) in $\sigma_4$. The detail of this parametrization was firstly proposed by \cite{DV16}.

\subsection{Solar Arrays}
The two symmetrical solar arrays can be modeled with the block \texttt{\textbf{1 port flexible body}}  available in SDTlib.
This block:
\begin{itemize}
\item computes the dynamic model $\left[\mathbf M_{P_1 }^{{\mathcal{A}}_1} \right]_{{\mathcal{R}}_{a_1 } } ({\mathrm{s}})$ of the body ${\mathcal{A}}_1$ at the point $P_1$ and projected in the body frame ${\mathcal{R}}_{a_1 }$, that is the matrix transfer from the acceleration twist $[{\ddot{\mathbf x} }_{P_1 } ]_{{\mathcal{R}}_{a_1 } }$ imposed at point $P_1$ by the main body $\mathcal{B}$ to the reaction wrench $\left[\mathbf W_{\mathcal{B}/{\mathcal{A}}_1 ,P_1} \right]_{{\mathcal{R}}_{a_1}}$ applied by body $\mathcal{B}$ on the body ${\mathcal{A}}_1$,
\item and applies  the action/reaction principle: 
\begin{equation}
	\left[\mathbf W_{{\mathcal{A}}_1 /\mathcal{B},P_1 }\right]_{{\mathcal{R}}_{a_1 } } =-\left[\mathbf W_{\mathcal{B}/{\mathcal{A}}_1 ,P_1 }\right]_{{\mathcal{R}}_{a_1 } }
\end{equation}
\end{itemize}

Assuming the damping ratios are null ($\xi =0$), one can express the model as a $6\times 6$ transfer matrix:
\begin{equation}
	\mathbf M_{P_1 }^{{\mathcal{A}}_1 } ({\mathrm{s}})=\mathbf D_{P_1 0}^{{\mathcal{A}}_1 } +\sum_{j=1}^{n=3} \frac{\mathbf l_{j,P_1 }^T \mathbf l_{j,P_1 } \omega_j^2 }{{{\mathrm{s}}}^2 +\omega_j^2 }
	\label{eq:flexapp}
\end{equation}

where $\omega_j$ ($j=1,2,3$) are the frequencies of three flexible modes and $\mathbf l_{j,P_1 }$ is the $1\times 6$ modal participation factor vector at the connection point $P_1$ of the $j$-th solar array flexible mode.
$\mathbf D_{P_1 0}^{{\mathcal{A}}_1 }$ is finally the $6\times 6$ residual mass "rigidly" attached to the point $P_1$. 
Note that Eq. \ref{eq:flexapp} is intrinsic and can be projected in any frame.

The low-frequency (DC) gain of the model $\mathbf M_{P_1 }^{{\mathcal{A}}_1 } (0)$ is always equal to the total mass matrix $\mathbf D_{P_1 }^{{\mathcal{A}}_1 }$ of the body ${\mathcal{A}}_1$ at the point $P_1$:
\begin{equation}
	\mathbf M_{P_1}^{{\mathcal{A}}_1 } (0)=\mathbf D_{P_1}^{{\mathcal{A}}_1 } ={\bm{\tau}}_{A_1 P_1 }^T {\underbrace{\left[ \begin{array}{cc}
				m^{{\mathcal{A}}_1 } \mathbf{I}_3  & 0_{3\times 3} \\
				0_{3\times 3}  & \mathbf J_{A_1 }^{{\mathcal{A}}_1 } 
			\end{array}\right]} }_{\mathbf D_{A_1 }^{{\mathcal{A}}_1 } } {\bm{\tau }}_{A_1 P_1}
\end{equation}
where ${\bm{\tau}}_{A_1 P_1 }$ is the kinematic model between points $A_1$ and $P_1$.
As consequence:
\begin{equation}
	\mathbf D_{P_1 0}^{{\mathcal{A}}_1 } = \mathbf D_{P_1 }^{{\mathcal{A}}_1 } -\sum_{j=1}^{n=3} \mathbf l_{j,P_1 }^T \mathbf l_{j,P_1 }
\end{equation} 

Since it is possible to declare the mass $m^{{\mathcal{A}}_1 }$, the components of the inertia tensor $\mathbf J_{A_1 }^{{\mathcal{A}}_1 }$, the 3 components of the vectors $\overrightarrow{O_{a_1}A_1}$,  $\overrightarrow{O_{a_1} P_1 }$ and  all the components of the modal participation factors $\mathbf l_{j,P_1 }$as uncertain parameters (\texttt{\textbf{ureal}}), the block \texttt{\textbf{1 port flexible body }}includes a WC analysis to check that the residual mass $\mathbf D_{P_1 0}^{{\mathcal{A}}_1 }$ is always a definite positive matrix for any parametric configurations.

\subsection{Spacecraft Assembly}

\begin{figure}[ht!]
    \centering
    \includegraphics[width=\columnwidth]{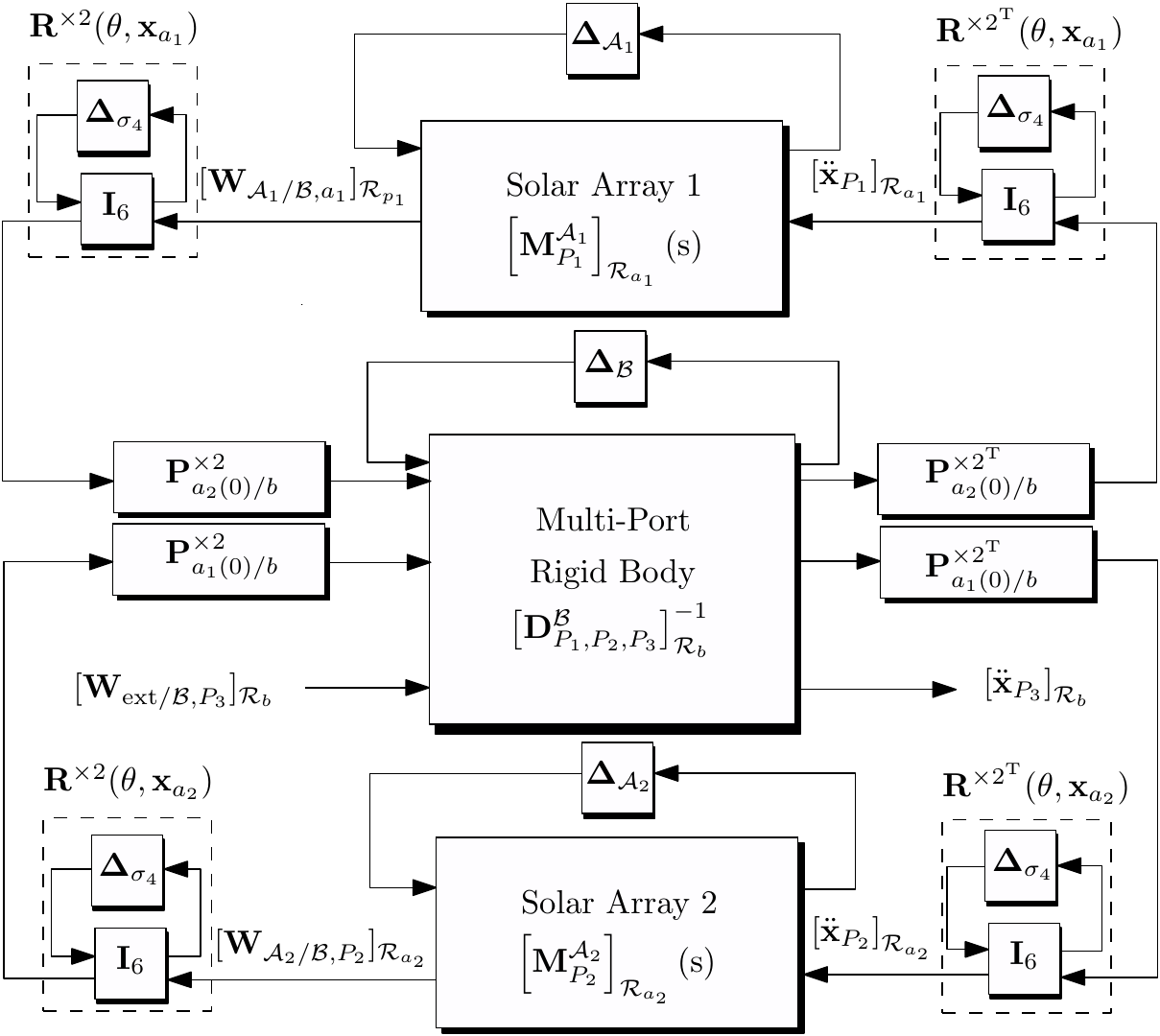}
    \caption{Spacecraft LFT assembly.}
    \label{fig:assembly_sdt}
\end{figure}

\begin{figure}[ht!]
    \centering
    \includegraphics[width=1\columnwidth]{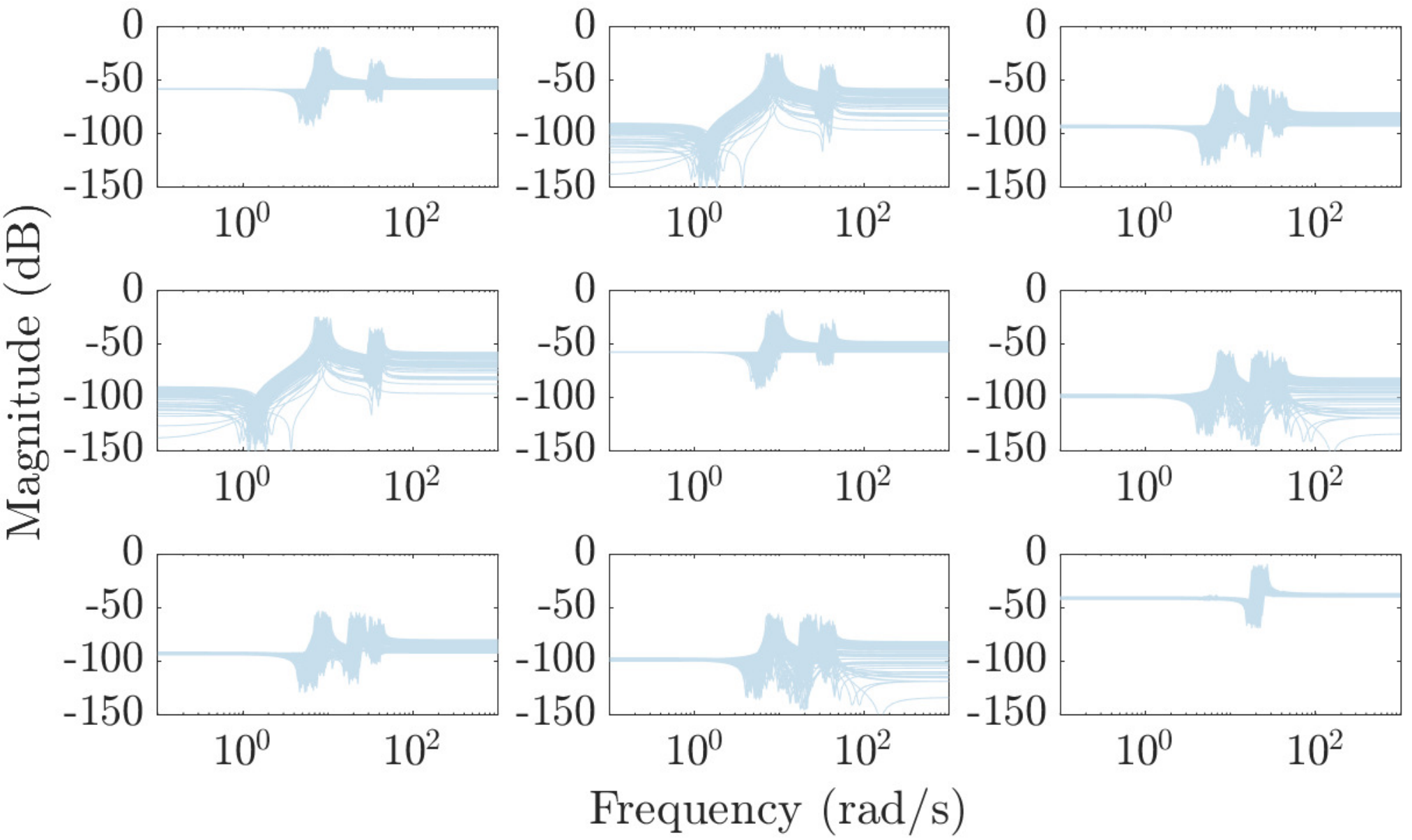}
    \caption{$3\times 3$ open-loop transfer function among external torques applied to the central body CoM and its angular accelerations.}
    \label{fig:open_loop_sigma}
\end{figure}

The assembled spacecraft model is depicted in Fig. \ref{fig:assembly_sdt}, where the previously described sub-systems can be distinguished. In the same diagram the blocks $\bm{\Delta}_\bullet$ represent the parametric uncertainties and varying parameters of the various sub-systems in an upper LFT form:
\begin{itemize}
	\item $\bm{\Delta}_\mathcal{B}=\mathrm{blkdiag}\left(\delta_{m^\mathcal
	B}\mathbf{I}_3,\,\delta_{_{xx}\mathbf{J}^\mathcal{B}_B},\,\delta_{_{yy}\mathbf{J}^\mathcal{B}_B},\,\delta_{_{zz}\mathbf{J}^\mathcal{B}_B}\right)$ incorporates the parametric uncertainties on the mass and diagonal terms of the main body inertia matrix,
	\item $\bm{\Delta}_\mathcal{A_\bullet}=\mathrm{blkdiag}\left(\delta_{\omega_1}\mathbf{I}_4,\delta_{\omega_2}\mathbf{I}_4,\delta_{\omega_3}\mathbf{I}_4\right)$ incorporates the parametric uncertainty on the value of frequency of the first three modes of the solar panels,
	\item $\bm{\Delta}_{\sigma_4}= \sigma_4 \mathbf{I}_8$ expresses the variations of the solar arrays' rotation angle.
\end{itemize}
The values of each of these uncertainties are specified in Table \ref{tab:data}.
Figure \ref{fig:open_loop_sigma} shows the singular values of the $3\times 3$ transfer function among external torques $\left[\mathbf{W}_{\mathrm{ext}/\mathcal{B},P_3}\right]_{\mathcal{R}_b}\left\{4:6\right\}$ applied to the central body CoM and its angular accelerations $\left[\ddot{\mathbf{x}}_{P_3}\right]_{\mathcal{R}_b}\left\{4:6\right\}$. Note how the system uncertainties determine a big shift of flexible mode frequencies.

\begin{table*}
    \renewcommand{\arraystretch}{1.3}
	\label{tab:spacecraft_data_sadm}
	\centering
	\begin{footnotesize}
		\resizebox{\textwidth}{!}{\begin{tabular}{|cllrr|}
				\hline
				\textbf{System} & \textbf{Parameter} & \textbf{Description} & \textbf{Nominal Value}  & \textbf{Uncertainty}\\ \hline
				& $m^{\mathcal{B}}$ & Mass & $1000\,\mathrm{kg}$ & $\pm 20\%$\\
			&  $\left[\mathbf{J}_B^\mathcal{B}\right]_{\mathcal{R}_{b}}$ & Inertia in $\mathcal{R}_{b}$ frame w.r.t. $B$&  $\left[\begin{array}{ccc}
				75 & 1 & 2 \\
				& 40 & -1 \\
				sym & & 80
			\end{array}\right]\mathrm{kg\, m^2}$ & $\left[\begin{array}{ccc}
				\pm 20\% & 0 & 0 \\
				& \pm 20\% & 0 \\
				sym & & \pm 20\%
			\end{array}\right]$\\
			\multirow{-3}{*}{\shortstack{Central \\ Body \\$\mathcal{B}$}}& $\left[\overrightarrow{O_bB}\right]_{\mathcal{R}_b}$ & CoM in $\mathcal{R}_{b}$ frame &  $\left[0.35\,\,1.50\,\,0.50\right]^\mathrm{T}\,\mathrm{m}$ & $-$ \\
				& $\left[\overrightarrow{O_bP_1}\right]_{\mathcal{R}_b}$ & Attachment $\mathcal{A}_1$  & $\left[0.4\,\,1.4\,\,1\right]^\mathrm{T}\,\mathrm{m}$ & $-$\\
				& $\left[\overrightarrow{O_bP_2}\right]_{\mathcal{R}_b}$ & Attachment $\mathcal{A}_2$  & $\left[0.4\,\,1.4\,\,0\right]^\mathrm{T}\,\mathrm{m}$ & $-$\\ \hline
				& $m^{\mathcal{A}_\bullet}$ & Mass & $43\,\mathrm{kg}$& $-$\\
				&  $\left[\mathbf{J}_{A_\bullet}^{\mathcal{A}_\bullet}\right]_{\mathcal{R}_{a_\bullet}}$ & Inertia in $\mathcal{R}_{a_\bullet}$ frame w.r.t. $A_\bullet$ &  $\left[\begin{array}{ccc}
					17& 0 & 0 \\
					& 62 & 0 \\
					sym & & 80
				\end{array}\right]\mathrm{kg\, m^2}$& $-$\\
				& $\left[\overrightarrow{O_{a_\bullet}A_\bullet}\right]_{\mathcal{R}_{a_\bullet}}$ & $\mathcal{A}_\bullet$ CG in $\mathcal{R}_{a_\bullet}$ frame &  $\left[2.07\,\,0\,\,0\right]^\mathrm{T}\,\mathrm{m}$& $-$\\
				& $\left[\omega_1^{\mathcal{A}_\bullet}\, \omega_2^{\mathcal{A}_\bullet}\,\omega_3^{\mathcal{A}_\bullet}\right]$ & Flexible modes' frequencies & $\left[5.6\,\,19.3\,\,35.4 \right]\,\mathrm{rad/s}$ &$\left[\pm 20\%\,\,\pm 20\%\,\,\pm 20\%\right]$ \\
				& $\xi_1^{\mathcal{A}_\bullet}, \xi_2^{\mathcal{A}_\bullet},\xi_3^{\mathcal{A}_\bullet}$ & Flexible modes' damping & $0.005$ & $-$\\
				\multirow{-6}{*}{\shortstack{Solar \\ Array $\mathcal{A}_\bullet$}} & $\mathbf{L}_{P_\bullet}^{\mathcal{A}_\bullet}$ & Modal participation factors & $\left[\begin{array}{cccccc}
					0 & 0 & -5.12 & 0 & 12.5 & 0\\
					0 & 0 & 0 & -3.84 & 0 & 0 \\
					0 & 0 & -2.97 & 0 & 2.51 & 0\\
				\end{array}\right]\,\mathrm{\sqrt{kg},m\sqrt{kg}}$ & $-$\\ 
				& $\sigma_4=\tan(\theta_\bullet/4)$ & Rotor angular position & 0 & $\left[-1;1\right]$\\ \hline
		\end{tabular}}
	\end{footnotesize}
	\caption{Spacecraft data.}
	\label{tab:data}
\end{table*}
 
\section{Control Synthesis}
\label{sec:control}

\subsection{Problem statement}

We consider the robust design of a $3$-axis  structured attitude control law to meet:
\begin{itemize}
\item (\texttt{\textbf{Req1}}) the pointing requirement, defined by the $3\times 1$ vector of Absolute Performance Error (\textbf{APE} $=\left[4\quad4 \quad20\right]^\mathrm{T}pi/180\cdot10^{-3}$ rad), in spite of low frequency orbital disturbances (characterized by the $3\times 1$ upper bound on the magnitude $\mathbf{T}_\mathrm{ext}=\left[0.03\quad0.01\quad 0.02\right]^\mathrm{T}$ Nm), 
\item (\texttt{\textbf{Req2}}) stability margins characterized by an upper bound $\gamma$ on the $H_{\infty }$-norm of the input sensitivity function,
\end{itemize}
while minimizing (\texttt{\textbf{Req3}}) the variance on the torque applied by the reaction wheel system on the spacecraft in response to the star sensor and gyrometer noises characterized by their Power Spectral Density (PSD), respectively \textbf{PSD}$^\mathrm{SST}=10^{-8}\mathbf{I}_3\,\mathrm{rad^2 s}$ and \textbf{PSD}$^\mathrm{GYRO}=10^{-10}\mathbf{I}_3\,\mathrm{rad^2/s}$ (assumed to be equal for the $3$ components).

The value $\gamma =1.5$ ensures on each of the 3 axes:
\begin{itemize}
\item a disk margin $>1/\gamma =0.667$,
\item a gain margin $>\frac{\gamma }{\gamma -1}=3\,(9.542\,\mathrm{dB})$,
\item a phase margin $>2\arcsin \frac{1}{2\gamma }=38.9\,\mathrm{deg}$. 
\end{itemize}

The requirements \texttt{\textbf{Req1}} and \texttt{\textbf{Req2}} must be met for any values of the uncertain mechanical parameters regrouped in the block $\bm{\Delta}=\mathrm{blkdiag}(\bm{\Delta}_{\mathcal{A}_1},\bm{\Delta}_{\mathcal{B}},\bm{\Delta}_{\mathcal{A}_2})$ and geometrical configuration $\theta$ of the solar arrays, expressed in the uncertain block $\bm{\Delta}_{\sigma_4}^T = \mathrm{blkdiag}(\bm{\Delta}_{\sigma_4},\bm{\Delta}_{\sigma_4},\bm{\Delta}_{\sigma_4},\bm{\Delta}_{\sigma_4})$.

Three reaction wheels connected to the central body CoM and aligned with the spacecraft reference body axes are used by the ACS. Their dynamics is  modeled as a second-order low-pass filter with a natural frequency of 100 Hz and natural damping equal to 0.7:
\begin{equation}
	\mathrm{\textbf{RW}}(\mathrm s) = \frac{(200\pi)^2}{\mathrm s^2 + 1.4\cdot200\pi \mathrm s + (200\pi)^2}\mathbf{I}_3
\end{equation}
The star sensor is modeled with a first-order low-pass filter with cut-off frequency of 8 Hz:
\begin{equation}
	\textbf{SST}(\mathrm{s}) = \frac{16\pi}{\mathrm s+16\pi}\mathbf{I}_3
\end{equation}

The gyrometer is modeled as a first-order too with a cut-off frequency of 200 Hz:
\begin{equation}
	\textbf{GYRO}(\mathrm{s}) = \frac{400\pi}{\mathrm s+400\pi}\mathbf{I}_3	
\end{equation}

The control-loop is characterized by a delay of 10 ms, that is modeled as 2nd-order Pade approximation \textbf{DELAY}(s).

\begin{figure*}[th!]
    \centering
    \includegraphics[width=.9\linewidth]{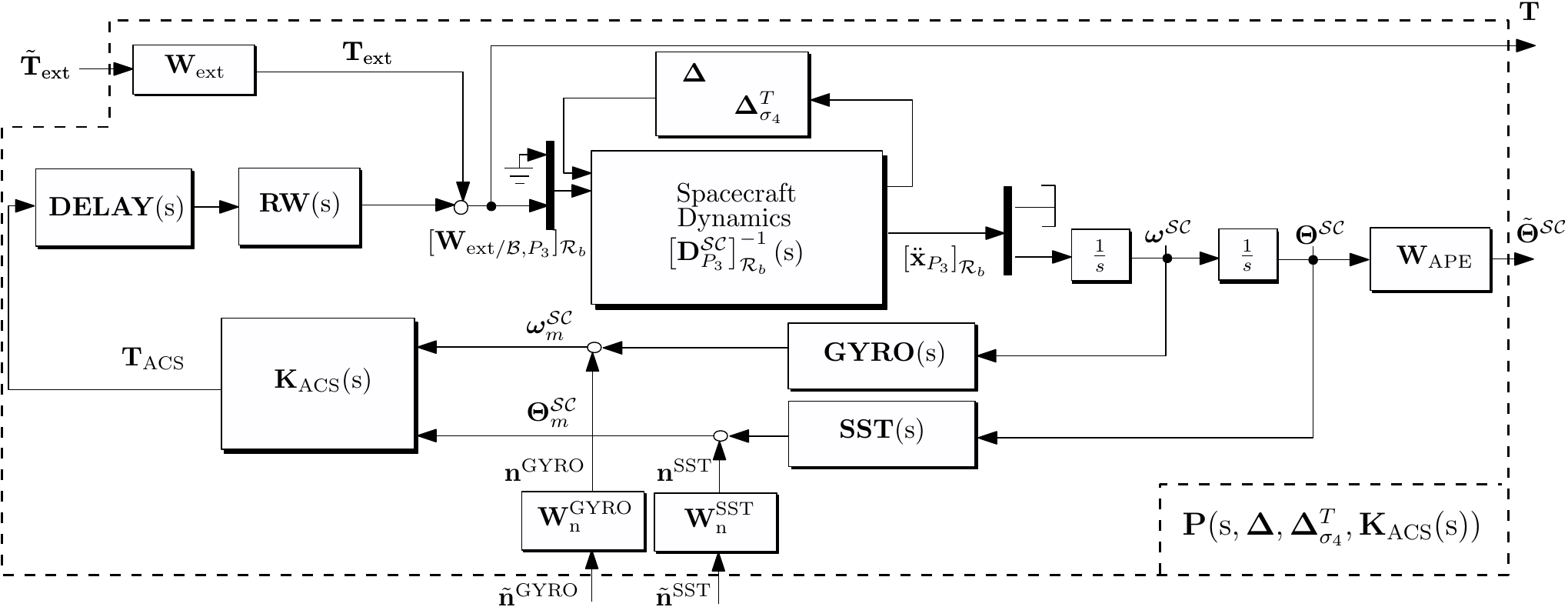}
    \caption{Generalized plant for robust control synthesis.}
    \label{fig:AOCS}
\end{figure*}

The closed-loop generalized plant $\mathbf{P}(\mathrm{s},\bm{\Delta},\bm{\Delta}_{\sigma_4}^T,\mathbf{K}_\mathrm{ACS}(\mathrm{s}))$ used for the robust control synthesis is shown in Fig. \ref{fig:AOCS}. The LFT model of the spacecraft dynamics presented in Fig. \ref{fig:assembly_sdt} is now condensed in the upper LFT $\mathcal{F}_u\left(\left[\mathbf{D}_{P_3}^\mathcal{SC}\right]^{-1}_{\mathcal{R}_b}(\mathrm s),\mathrm{blkdiag}(\bm{\Delta},\bm{\Delta}^T_{\sigma_4})\right)$. The following weighting filters are used to normalize the inputs and outputs: $\mathbf{W}_\mathrm{ext}=\mathbf{T}_\mathrm{ext}$, $\mathbf{W}^\mathrm{SST}_n = \sqrt{\textbf{PSD}^\mathrm{SST}}$, $\mathbf{W}^\mathrm{GYRO}_n = \sqrt{\textbf{PSD}^\mathrm{GYRO}}$, $\mathbf{W}_\mathrm{APE}=(\mathrm{diag}(\textbf{APE}))^{-1}$.

\begin{figure}[ht!]
    \centering
    \includegraphics[width=\columnwidth]{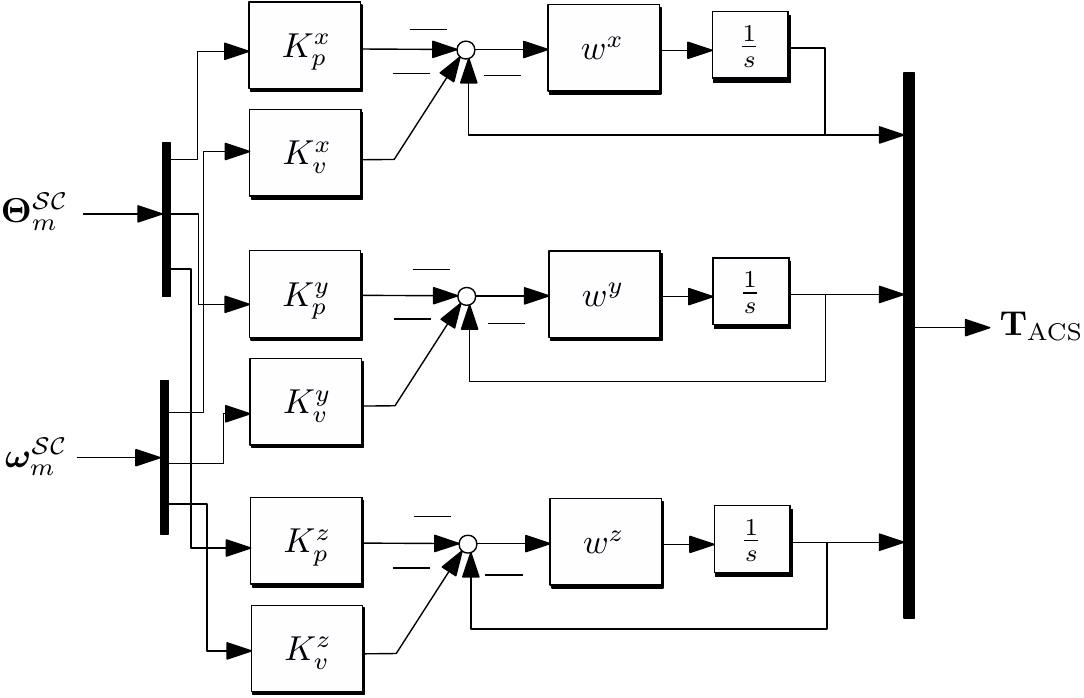}
    \caption{Structured controller $\mathbf{K}_\mathrm{ACS}(\mathrm{s})$.}
    \label{fig:AOCS_block}
\end{figure}

Finally the block $\mathbf{K}_\mathrm{ACS}(\mathrm s)$ represents the structured $3\times 6$ attitude controller to be synthesized. The chosen structure is shown in Fig. \ref{fig:AOCS_block}. It is a decentralized controller composed of a proportional-derivative controller (the  gains $K_p^i$  and $K_v^i$, $i=x,y,z$) with a first order low pass filter (characterized by the cut-off frequency $w^i$, $i=x,y,z$) per axis. 

The set of the 9 controller tunable parameters ($K_p^i$, $w^i$  and $K_v^i$, $i=x,y,z$) are roughly initialized assuming a rigid 3-axis decoupled spacecraft and in order to:
\begin{itemize}
	\item reject a constant orbital disturbance $\mathbf{T}_\mathrm{ext}$  with a steady state pointing error $\bm{\Theta}^\mathrm{SC}$ lower than the $\textbf{APE}(i)$ requirement on each axis $i=x,y,z$,
	\item tune the 2-nd order closed-loop dynamics of each axis with a damping ratio $\xi=0.7$  and a given frequency bandwidth $\omega^i$.
\end{itemize}

Indeed, under these assumptions, the open-loop model between the control torque and the pointing error is equal to $\frac{1}{\mathbf{J}_{B}^{\mathcal{SC}}s^2}$, where the nominal $3\times 3$ inertia $\mathbf{J}_{B}^{\mathcal{SC}}$ on the whole spacecraft at point $B$ can be computed from the DC gain of the nominal model $\left[\mathbf{D}_{P_3}^\mathcal{SC}\right]^{-1}_{\mathcal{R}_b}(\mathrm s)$:
\begin{equation}
	\mathbf{J}_{B}^{\mathcal{SC}} =\left[ \left[\mathbf{D}_{P_3}^\mathcal{SC}\right]^{-1}_{\mathcal{R}_b}(0)\right]^{-1}\{4:6;4:6\}
\end{equation}

Then the tuning:
\begin{align}
	\begin{split}
		K_{p}^i &=\mathbf J_B^{\mathcal{SC}} \{i,i\}\,\omega_i^{r^2} \\
		K_{v}^i &=2\,\xi \,\mathbf J_B^{\mathcal{SC}} \{i,i\}\,\omega_i^r
	\end{split}
\end{align}
with $\omega_i^r$ required bandwidth on $i$-th axis ($i=1,2,3$), ensures the needed closed-loop dynamics and a disturbance rejection function expressed as:
\begin{equation}
	\frac{\bm{\Theta}^\mathcal{SC} \{i\}}{\mathbf T_\mathrm{ext} \{i\}}({\mathrm{s}})=\frac{1}{\mathbf J_B^{\mathcal{SC}} \{i,i\}}\frac{1}{{{\mathrm{s}}}^2 +2\,\xi \,\omega_i^r {\mathrm{s}}+\omega_i^{r^2} }
\end{equation}
Thus the required bandwidth $\omega_i^r$ to meet the absolute pointing error requirement in steady state ($\bm{\Theta}^\mathcal{SC} \{i\} \le \textbf{APE}\{i\}$) is:
\begin{equation}
	\omega_i^r \ge \sqrt{\frac{\mathbf T_\mathrm{ext} \{i\}}{\mathbf J_B^{\mathcal{SC}} \{i,i\}\textbf{APE}\{i\}}}
\end{equation}

The frequency of the first order low pass filter $\omega^i$ is tuned at $20\omega_i^r$ on each axis.

\begin{figure}
    \centering
    \includegraphics[width=\columnwidth]{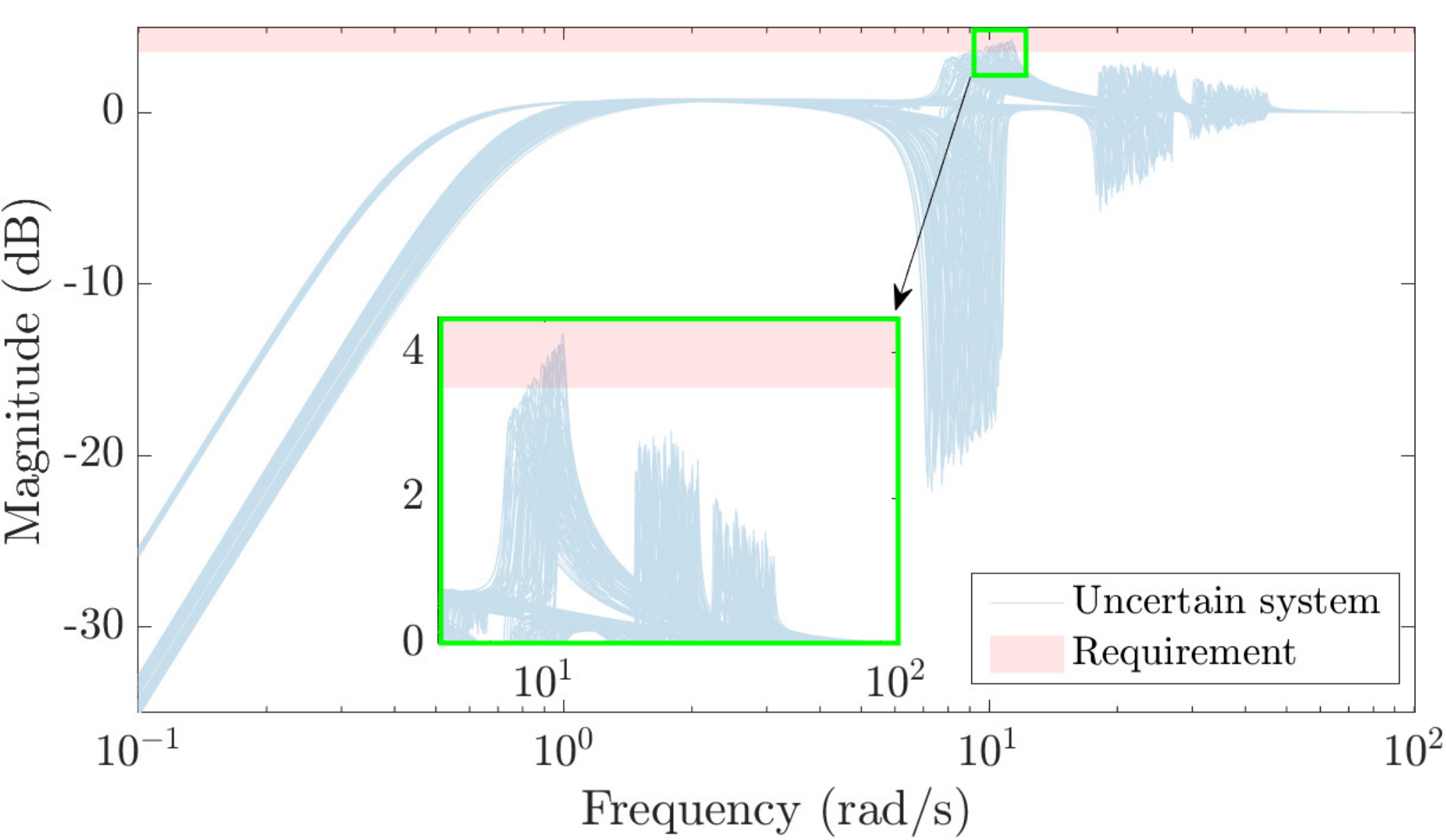}
    \caption{Singular Values of the transfer $\bar{\mathbf S} (\mathrm{s},\bm{\Delta},\bm{\Delta}^T_{\sigma_4},\widehat{\mathbf K})$ with the initial guess controller.}
    \label{fig:initialGuess}
\end{figure}

This initial tuning, based on simplified assumptions (that does not verify the stability requirement (\texttt{\textbf{Req2}}) for some parametric configurations as shown in Fig. \ref{fig:initialGuess} and does not minimize (\texttt{\textbf{Req3}})), is useful to initialize the non-convex optimization problem presented in the next section. 

\subsection{Robust control synthesis}

In this section a robust controller is synthesized thanks to the \texttt{systune} routine available in MATLAB, based on the non-convex optimization algorithm proposed by \cite{apkarian2015parametric}. The three performance criteria can be easily translated into the following $\mathcal{H}_2/\mathcal{H}_\infty$ problem.
The optimization function to be optimized, that translates (\texttt{\textbf{Req3}})), is:
\begin{equation}
	\widehat{J} =\max_{\bm{\Delta },\bm{\Delta}^T_{\sigma_4}} \left|\left|\mathbf P_{\tilde{\mathbf n} \to \mathbf T} ({\mathrm{s}},\bm{\Delta },\bm{\Delta}^T_{\sigma_4} ,\widehat{\mathbf K} )\right|\right|_2 
\end{equation}
with
\begin{equation}
	 \widehat{\mathbf K} =\arg \min_{\mathbf K} \max_{\bm{\Delta },\bm{\Delta}^T_{\sigma_4}} \left|\left|\mathbf P_{\tilde{\mathbf n} \to \mathbf T} ({\mathrm{s}},\bm{\Delta },\bm{\Delta}^T_{\sigma_4},\mathbf K)\right|\right|_2
\end{equation}
Here $\mathbf{K}$ represents the set of 9 control parameters to be tuned; $\left|\left|\mathbf P_{\tilde{\mathbf n} \to \mathbf T} ({\mathrm{s}},\bm{\Delta },\bm{\Delta}^T_{\sigma_4},\mathbf K)\right|\right|_2$ represents the $\mathcal{H}_2$-norm of the multi-input multi-output (MIMO) transfer from the normalized vector noise $\tilde{\mathbf{n}}=\left[\tilde{\mathbf{n}}^{\text{GYRO}^\mathrm{T}}\,\, \tilde{\mathbf{n}}^{\text{SST}^\mathrm{T}}\right]^\mathrm{T}$ to the control torque vector $\mathbf T$ in the closed-loop generalized plant presented in Fig. \ref{fig:AOCS}. 

The pointing requirement (\texttt{\textbf{Req1}}) corresponds to the \textit{hard constraint} on the $\mathcal{H}_\infty$-norm of the MIMO transfer from the normalized external orbital perturbations $\tilde{\mathbf{T}}_\text{ext}$ to normalized pointing performance $\tilde{\mathbf \Theta}$:
\begin{equation}
	\max_{\bm{\Delta},\bm{\Delta}^T_{\sigma_4}}\left|\left| \mathbf P_{{\tilde{\mathbf T} }_\text{ext} \to \tilde{\bm{\Theta }} } ({\mathrm{s}},\bm{\Delta},\bm{\Delta}^T_{\sigma_4} ,\widehat{\mathbf K} )\right|\right|_{\infty } \le 1
\end{equation}
Finally the stability requirement (\texttt{\textbf{Req2}}) on the input sensitivity function $\bar{\mathbf{S}}({\mathrm{s}},\bm{\Delta},\bm{\Delta}^T_{\sigma_4} ,{\mathbf K} )=\mathbf P_{{\tilde{\mathbf T} }_\text{ext} \to \mathbf{T} } ({\mathrm{s}},\bm{\Delta},\bm{\Delta}^T_{\sigma_4} ,{\mathbf K})$ is translated into the \textit{hard constraint} on the $\mathcal{H}_\infty$-norm:
\begin{equation}
	\max_{\bm{\Delta},\bm{\Delta}^T_{\sigma_4}}\left|\left|\bar{\mathbf S} (\mathrm{s},\bm{\Delta},\bm{\Delta}^T_{\sigma_4},\widehat{\mathbf K} )\right|\right|_{\infty} \le \gamma
\end{equation}

The WC performance indexes obtained by running \texttt{systune} optimization are summarized in Table \ref{tab:systune}. Since all of them are lower than 1, all the control requirements are met for any parametric uncertainties and any geometric configurations according to the heuristic WC parametric and geometric configurations detected by \texttt{systune}.

\begin{table}[ht!]
    \centering
    \begin{tabular}{|c|c|c|}
    \hline
        \textbf{WC APE} & \textbf{WC Stability Margins} & \textbf{WC Variance}  \\ \hline
        0.9999 & 0.9999 & 0.1223 \\ \hline
    \end{tabular}
    \caption{Robust control synthesis WC performance indexes}
    \label{tab:systune}
\end{table}

The values of the tunable parameters of the controller are detailed in Table \ref{tab:opt_gains}. In this table the initial guess and the optimized values are compared.

\begin{table}[ht!]
	\centering
	\begin{tabular}{|ccc|}
		\hline
		& \textbf{initial}  & \textbf{optimal}  \\ \hline
		$K_p^x$ & 429.7183 & 430.9168 \\
		$K_v^x$ & 804.2055 & 819.1336 \\
		$\omega^x$ & 14.9615 & 13.1309 \\
		$K_p^y$ & 143.2394 & 143.2494 \\
		$K_v^y$ & 463.8107 & 541.9954\\
		$\omega^y$ & 8.6473 & 13.4353 \\
		$K_p^z$  & 57.2958 &  57.2995 \\
		$K_v^z$ & 113.6789 &  145.4465 \\
		$\omega^z$ & 14.1124 & 2.6361 \\
		\hline
	\end{tabular}
	\caption{Controller tunable parameters: initial guess and optimal solution provided by \texttt{systune}.}
	\label{tab:opt_gains}
\end{table}

\section{Robust Control Analysis}
\label{sec:analysis}

The last step of the end-to-end design process is the V\&V campaign in order to provide a guarantee of robust stability and performance.
In this section we propose some analyses run with the MATLAB routine \texttt{wcgain} of the Robust Control Toolbox and the routines available in the SMAC Toolbox developed by \cite{biannic2016smac} and the STOchastic Worst-case Analysis Toolbox (STOWAT), which was first introduced by \cite{ThRoBi19_acc} and \cite{BiRoBeBoPrGi21_ejc}.
The analyses proposed in this articles have to be considered as simple examples while researchers are challenged to use the proposed benchmark to highlight the performance and check the limits of their own V\&V algorithms.

The basis of the robust analysis is the structured singular value $\mu$ theory presented in \cite{zhou1998essentials}. Given an LFT model  $\mathcal{F}_u\left(\mathbf{M}(\mathrm s),{\bm{\Delta}}_M\right)$, where $\mathbf M(\mathrm{s})$ is a continuous-time stable and proper real-rational transfer function representing the nominal closed-loop system and ${\bm{\Delta}_M}$ is a block of uncertainties (parametric or complex), the structured singular value $\mu$ is defined as:

\begin{equation}
	\mu(\mathbf{M}(\mathrm s)) = \frac{1}{\min_{\bm{\Delta}_M}\left\{\bar{\sigma}(\bm{\Delta}_M), \det(\mathbf{I}-\mathbf{M}(\mathrm s)\bm{\Delta}_M)=0\right\}}
\end{equation}
where $\bar{\sigma}(\bm{\Delta}_M)$ denotes the maximum singular values of $\bm{\Delta}_M$. 
The exact computation of $\mu(\mathbf{M}(\mathrm s))$ is an NP hard problem (\cite{braatz1994computational}) and several algorithms for the computation of the upper and lower bounds of $\mu((\mathbf{M}(\mathrm s)))$ are available in literature.

\subsection{Pointing requirement}
\begin{figure}[ht!]
    \centering
    \includegraphics[width=\columnwidth]{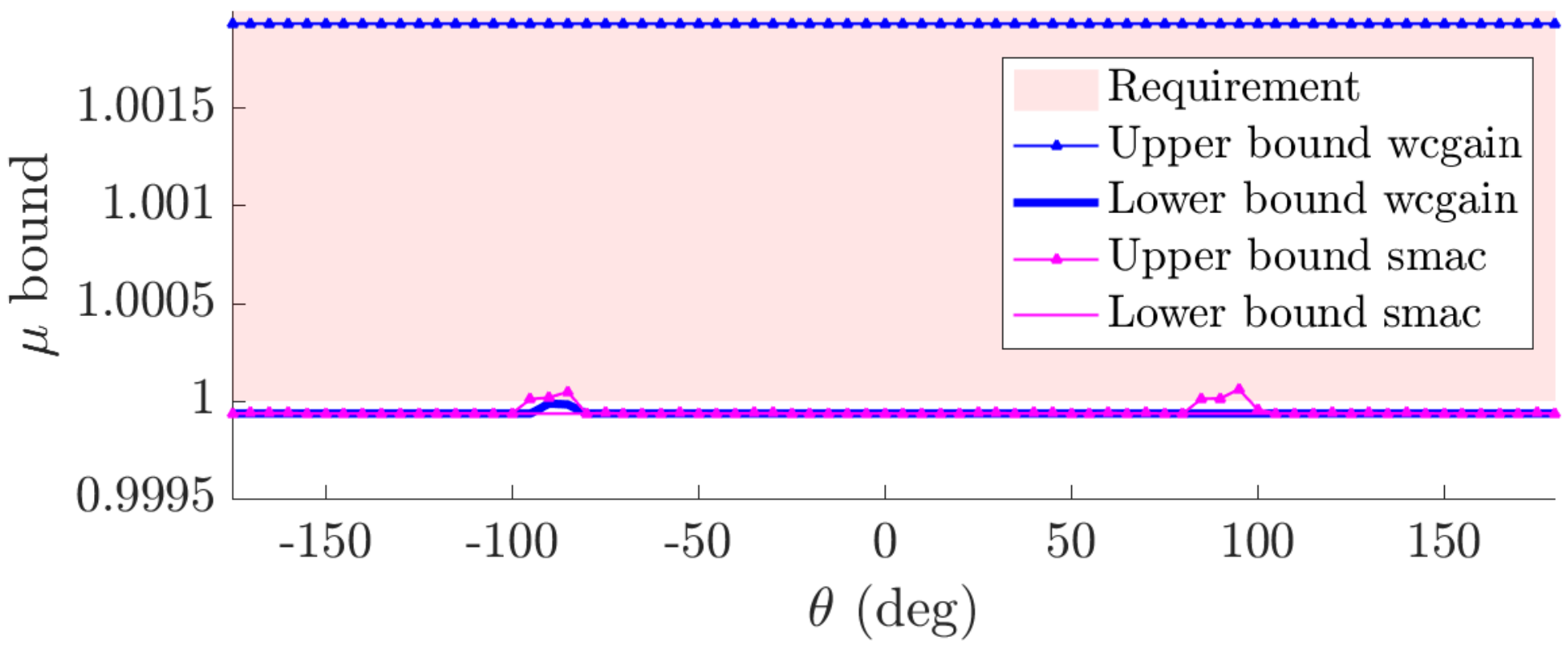}
    \caption{$\mu$ bounds for the pointing requirement constraint for a gridding of 72 different values of $\sigma_4$.}
    \label{fig:WC_APE}
\end{figure}

\begin{figure}[ht!]
	\centering
	\includegraphics[width=\columnwidth]{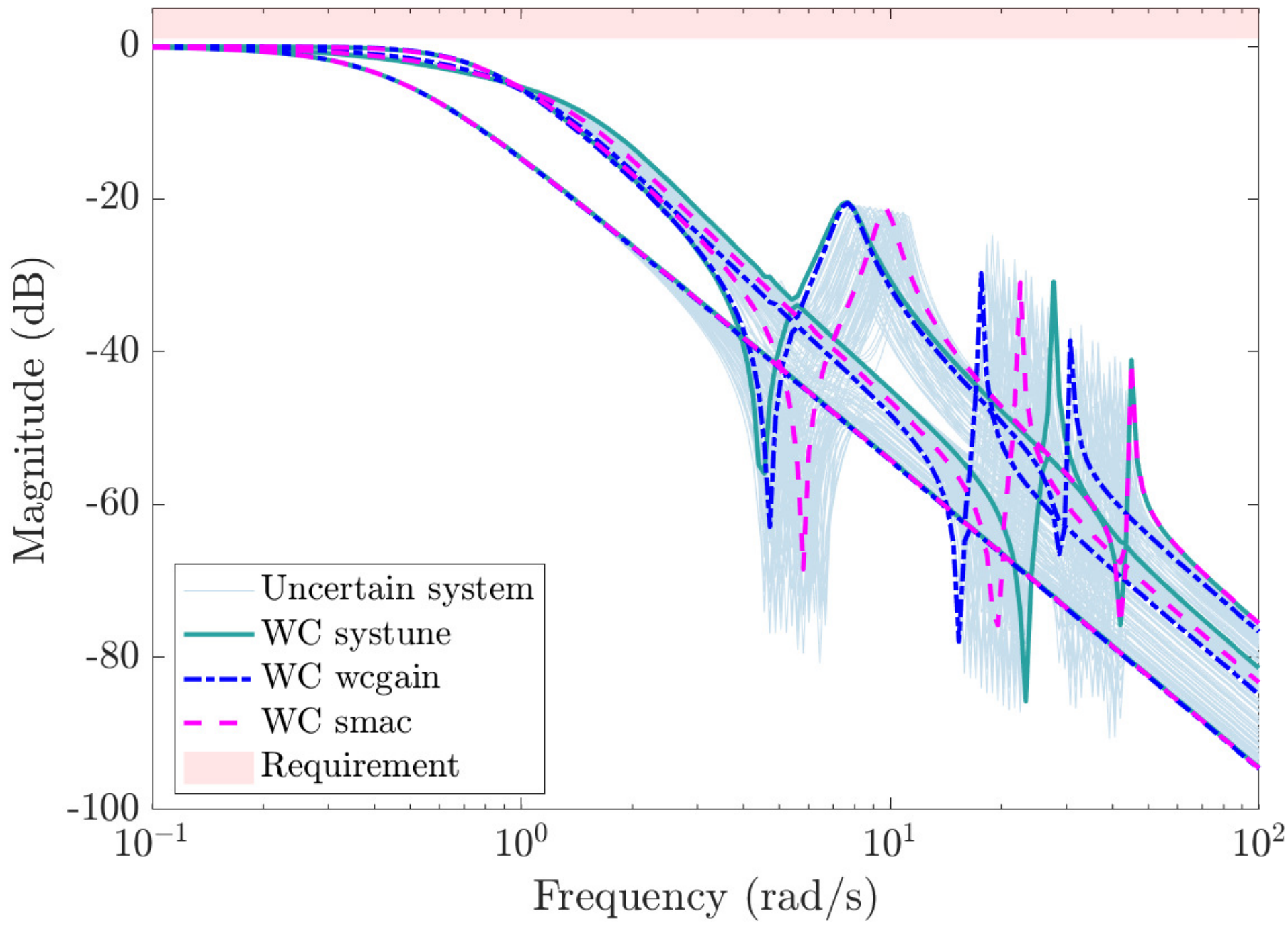}
	\caption{Singular Values of the transfer $\mathbf P_{{\tilde{\mathbf T} }_\text{ext} \to \tilde{\bm{\Theta }} } ({\mathrm{s}},\bm{\Delta},\bm{\Delta}^T_{\sigma_4} ,\widehat{\mathbf K})$.}
	\label{fig:sigma_WC_APE}
\end{figure}

We analyze in this section the robustness of the pointing requirement (\texttt{\textbf{Req1}}). The routine \texttt{wcgain} was found to fail in the computation of the upper and lower $\mu$ bounds due to the high number of repetitions of the parameter $\sigma_4$ in the uncertain block $\bm{\Delta}_{\sigma_4}^T$ equal to 32. However by using the routines \texttt{mulb} and \texttt{muub} (by restricting the analysis to the frequency bandwidth $\left[0\,,2\right]$ rad/s) available in the SMAC toolbox this computation brings respectively in 2.46 s and 5.54 s to a lower bound:
 $\underline{\mu}\left(\mathbf P_{{\tilde{\mathbf T} }_\text{ext} \to \tilde{\bm{\Theta }} } ({\mathrm{s}},\bm{\Delta},\bm{\Delta}^T_{\sigma_4} ,\widehat{\mathbf K} )\right) = 0.9999$ and an upper bound $\bar{\mu}\left(\mathbf P_{{\tilde{\mathbf T} }_\text{ext} \to \tilde{\bm{\Theta }} } ({\mathrm{s}},\bm{\Delta},\bm{\Delta}^T_{\sigma_4} ,\widehat{\mathbf K} )\right) = 1.2092$, which provides some conservatism due to the difference between the two bounds.
 Another analysis is done by sampling the parameter $\sigma_4$ over a grid with 72 points (corresponding to every 5 degrees between -175$^\circ$ and 180$^\circ$) and run the analysis by considering only the set of uncertainties $\bm{\Delta}$. The results obtained with \texttt{wcgain} (in 302.18 s with the option LMI) and SMAC (the lower bound in 31.39 s and the upper bound with LMI option on the frequency bandwidth $\left[0\,,2\right]$ rad/s in 201.31 s) are shown in Fig. \ref{fig:WC_APE}. The  bounds are almost independent of the solar array configuration. No WC parametric configuration violating the pointing requirement was isolated (the  lower bound is always under 1). On the frequency-response in Fig. \ref{fig:sigma_WC_APE} it can be also verified that the pointing requirement is actually saturated in low frequency for all parametric configurations.
 
\subsection{Stability margin requirement}

\begin{figure}[ht!]
    \centering
    \includegraphics[width=\columnwidth]{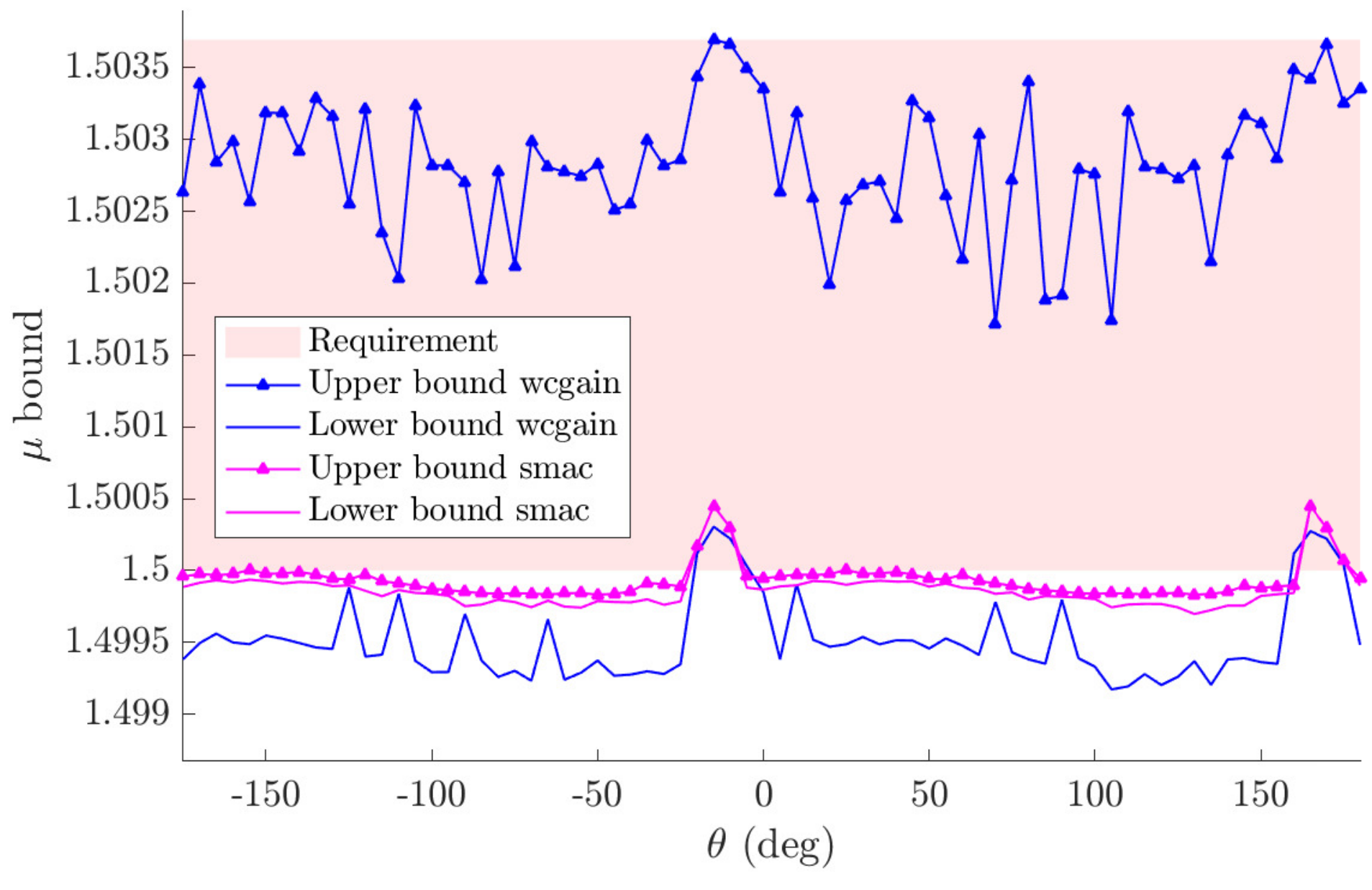}
    \caption{$\mu$ bounds for the stability requirement constraint for a gridding of 72 different values of $\sigma_4$.}
    \label{fig:WC_SENS}
\end{figure}

For the stability margin requirement (\texttt{\textbf{Req2}}) both \texttt{wcgain} and SMAC cannot provide the $\mu$ upper bound if the entire set of uncertainties is considered. The same grid of $\sigma_4$ with 72 values is then used to evaluate the $\mu$ bounds for each sampled system. The results obtained both with \texttt{wcgain} and SMAC are shown in Fig. \ref{fig:WC_SENS} and the computed WC are summarized in Table \ref{tab:mu_perf_sens}. Note that exactly the same options as for the validation of the pointing requirement are used for the two different approaches except for the frequency range used for the computation of the upper bound with SMAC toolbox that is now $[0\,,12]\,\mathrm{rad/s}$.
The following conclusions can be inferred:
\begin{itemize}
    \item There exists some critical cases (for which $\mu>1.5$) that were not detected by \texttt{systune} heuristic search even if the requirement is very slightly exceeded;
    \item Conservatism (difference between lower and upper bounds) obtained with SMAC toolbox is smaller than \texttt{wcgain};
    \item The WC configuration is almost the same one for SMAC and \texttt{wcgain} and corresponds roughly to the minimum values of the inertia of the central body. This results makes physical sense since in this configuration the impact of the flexible modes of the solar panels is more important;
    \item The total computational cost of SMAC toolbox (lower + upper $\mu$ bound) is $\approx 7$ times less than the one needed by \texttt{wcgain}. 
\end{itemize}

\begin{figure}[ht!]
    \centering
    \includegraphics[width=\columnwidth]{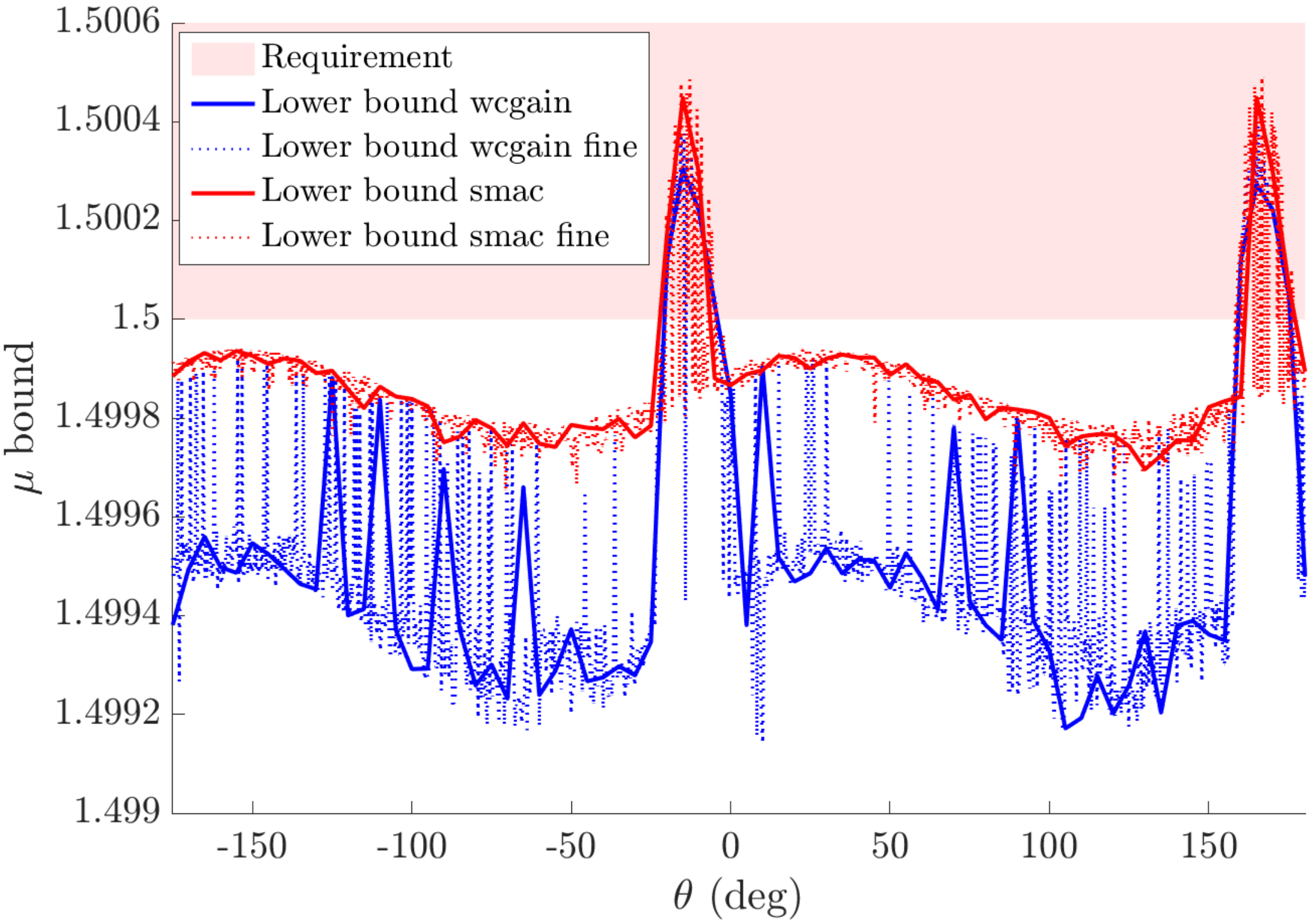}
    \caption{$\mu$ lower bounds for the stability requirement constraint for a gridding of 72 (solid lines) and 1001 (dotted lines) different values of $\sigma_4$.}
    \label{fig:WC_SENS_fine}
\end{figure}

\begin{figure}[ht!]
    \centering
    \includegraphics[width=\columnwidth]{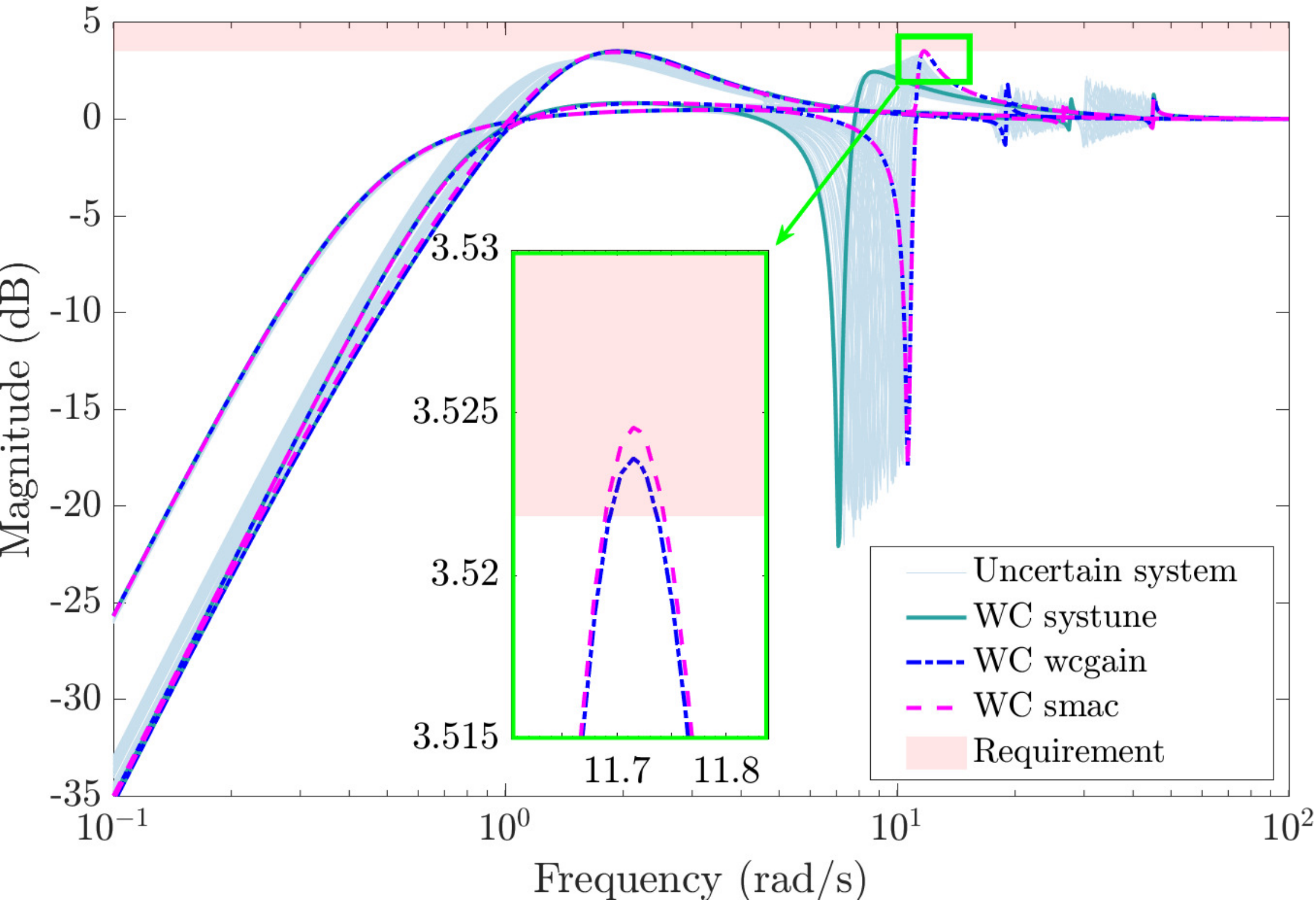}
    \caption{Singular Values of the transfer $\bar{\mathbf S} (\mathrm{s},\bm{\Delta},\bm{\Delta}^T_{\sigma_4},\widehat{\mathbf K} )$.}
    \label{fig:sigma_WC_SENS}
\end{figure}

A finer grid (1001 values of $\sigma_4$) is used to check better the $\mu$ lower bound. The result is shown in Fig. \ref{fig:WC_SENS_fine}. Following the figure a way longer analysis confirms the results obtained with the rough grid.
Figure \ref{fig:sigma_WC_SENS} shows the singular values of the sensitivity function $\bar{\mathbf S} (\mathrm{s},\bm{\Delta},\bm{\Delta}^T_{\sigma_4},\widehat{\mathbf K} )$ with an highlight on the computed WC configurations.

\begin{table*}[th!]
    \centering
    \resizebox{\textwidth}{!}{\begin{tabular}{|l|c|c|c|*{7}{c}|c|}
    \hline
     &  & & & \multicolumn{7}{c|}{\textbf{WC Parameters}}& \textbf{CPU} \\
     & \textbf{WC} $\underline{\mu}$  & \textbf{WC Frequency} & \textbf{WC} $\theta$ & $m^\mathcal{B}$ & $\mathbf{J}_B^\mathcal{B}\{1,1\}$ & $\mathbf{J}_B^\mathcal{B}\{2,2\}$ & $\mathbf{J}_B^\mathcal{B}\{3,3\}$ & $\omega_1^{\mathcal{A}_\bullet}$ & $\omega_2^{\mathcal{A}_\bullet}$ & $\omega_3^{\mathcal{A}_\bullet}$ & \textbf{Time}\\
     & &  (rad/s) & (deg) & (kg) & ($\mathrm{kg\,m^2}$) & ($\mathrm{kg\,m^2}$) & ($\mathrm{kg\,m^2}$) & (rad/s) & (rad/s) & (rad/s) & (min) \\
     \hline
     \texttt{wcgain} & 1.5003 & 11.7146 & -15 & 814.99 & 60 & 32.82 & 64 & 6.720 & 15.877 & 42.480 & 48.10 \\ 
     SMAC & 1.5004 & 11.7211 & -15 & 800 & 60 & 32 & 67.22 & 6.720 & 22.383 & 42.480 & 6.53 \\ \hline
    \end{tabular}}
    \caption{Stability margin performance WC results. CPU time accounts for lower and upper bound computation.}
    \label{tab:mu_perf_sens}
\end{table*}

\begin{figure*}[t!]
    \centering
    \includegraphics[width=.92\linewidth]{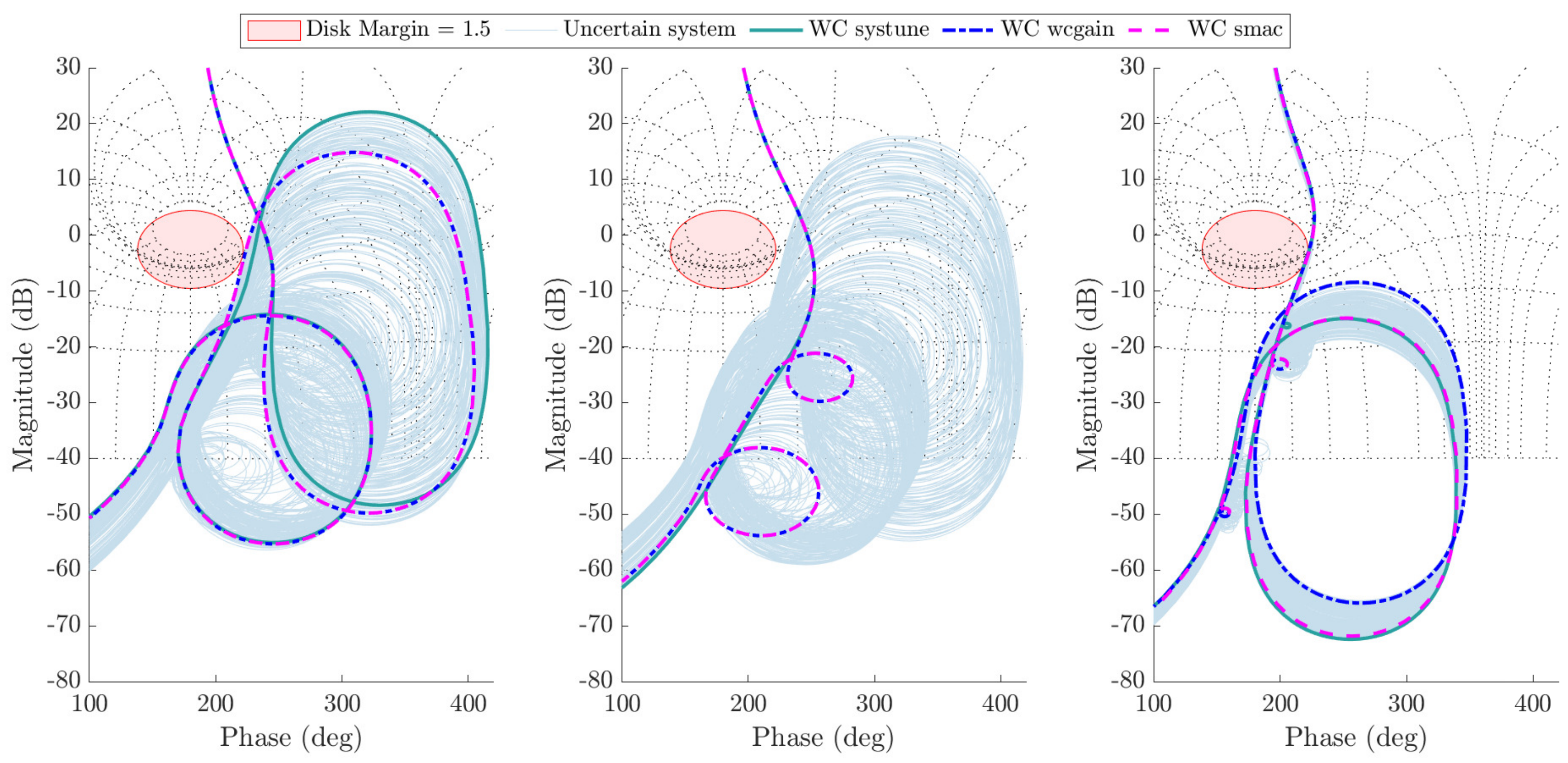}
    \caption{Nichols plot of open-loop transfer obtained by opening one control axis (x,y,z) at the time.}
    \label{fig:nychols}
\end{figure*}

For illustration of the correspondence of stability requirement with the required disk margin of $1/\gamma = 0.667$, Fig. \ref{fig:nychols} shows the Nichols plot obtained by opening one control axis (x,y,z) at the time. 

\subsubsection{Probabilistic stability margin requirement}
The STOWAT, is devoted to probabilistic $\mu$-analysis and is used in this work to perform probabilistic $\mathcal H_{\infty}$ performance analyses, to study \texttt{\textbf{Req2}} in more depth. The probabilistic $\mathcal H_{\infty}$ performance algorithm of the STOWAT is limited to Single Input Single Output (SISO) system analysis. Therefore, a loop-at-a-time analysis is performed for the considered MIMO model by taking a uniform probability distribution for each uncertainty. Similar to the \texttt{wcgain} and SMAC routines, the STOWAT computations should be performed on a reduced frequency range. The robust controller is first studied on the same range as used for the SMAC routines, $\left[0,12\right]$ rad/s. On this domain, it can be shown that the probability of non-performance is guaranteed to be less than $2.5\%$. The total CPU time for the three SISO analyses and the preliminary stability analysis is $\sim37$ minutes. More accurate, lower probabilities of performance violation can be computed at the cost of an increased CPU time. Using the previously acquired knowledge of the system, regarding the WC configurations presented in Fig. \ref{fig:sigma_WC_SENS}, a more efficient analysis can be done. This by considering only a small range focused on the detected WC configurations, $\left[11, 12\right]$ rad/s. Indeed, on this range, it can be computed in $\sim12$ minutes that the probability of non-performance is guaranteed to be less than $0.5\%$ for all three considered loops. For comparative purposes, the initial guess controller is analyzed at this same narrow frequency range. As expected, this controller performs worse. The probability of non-performance is guaranteed to be higher than that of the robust controller. Even tough, the probability of non-performance for the first and last SISO transfer is less than $0.5\%$, the probability for the second transfer indeed lies between $0.5\%$ and $3.7\%$.


\subsection{Variance requirement}
For the V\&V on the variance requirement (\texttt{\textbf{Req3}}) no existing tool to the authors' knowledge is able to provide an analytically guaranteed performance.

\section{Conclusion}
\label{sec:conclusion}

A tutorial on modeling, robust control and analysis has been presented for the design of the attitude control system of a  flexible spacecraft. 
The message the authors care about is that with the LFT framework offered by proper modeling, synthesis and analysis tools, as shown in this tutorial, industrial practice, largely based on simplified SISO nominal control synthesis and time-consuming Monte Carlo simulations, could benefit from these techniques to: directly obtain a robust controller, by taking into account in the model all system uncertainties from the very beginning of the control design phase; save time in V\&V process by directly detect worst-case configurations in frequency domain that could escape to sampled-based Monte Carlo simulations in time domain.

\begin{ack}
The authors thank Clément Roos (ONERA) for his advice and support.
\end{ack}

\bibliography{ifacconf}             

\begin{thebibliography}{11}
\providecommand{\natexlab}[1]{#1}
\providecommand{\url}[1]{\texttt{#1}}
\providecommand{\urlprefix}{URL }
\expandafter\ifx\csname urlstyle\endcsname\relax
  \providecommand{\doi}[1]{doi:\discretionary{}{}{}#1}\else
  \providecommand{\doi}{doi:\discretionary{}{}{}\begingroup
  \urlstyle{rm}\Url}\fi

\bibitem[{Alazard et~al.(2015)Alazard, Perez, Cumer, and Loquen}]{titop}
Alazard, D., Perez, J.A., Cumer, C., and Loquen, T. (2015).
\newblock Two-input two-output port model for mechanical systems.
\newblock \emph{AIAA GNC Conference}.

\bibitem[{Alazard and Sanfedino(2020)}]{sdt}
Alazard, D. and Sanfedino, F. (2020).
\newblock Satellite dynamics toolbox for preliminary design phase.
\newblock \emph{43rd Annual AAS Guidance and Control Conf.}, 30, 1461--1472.

\bibitem[{Apkarian et~al.(2015)Apkarian, Dao, and
  Noll}]{apkarian2015parametric}
Apkarian, P., Dao, M.N., and Noll, D. (2015).
\newblock Parametric robust structured control design.
\newblock \emph{IEEE Trans. on Automatic Control}, 60(7), 1857--1869.

\bibitem[{Balas et~al.(2021)Balas, Chiang, Packard, and
  Safonov}]{RCT_user_guide}
Balas, G., Chiang, R., Packard, A., and Safonov, M. (2021).
\newblock Robust control toolbox 3 user’s guide.
\newblock Technical report, MATLAB.

\bibitem[{Biannic et~al.(2021)Biannic, Roos, Bennani, Boquet, Preda, and
  Girouart}]{BiRoBeBoPrGi21_ejc}
Biannic, J.M., Roos, C., Bennani, S., Boquet, F., Preda, V., and Girouart, B.
  (2021).
\newblock Advanced probabilistic $\mu$-analysis techniques for {AOCS}
  validation.
\newblock \emph{European Journal of Control}, 62, 120--129.

\bibitem[{Biannic et~al.(2016)Biannic, Burlion, Demourant, Ferreres, Hardier,
  Loquen, and Roos}]{biannic2016smac}
Biannic, J., Burlion, L., Demourant, F., Ferreres, G., Hardier, G., Loquen, T.,
  and Roos, C. (2016).
\newblock The smac toolbox: a collection of libraries for systems modeling,
  analysis and control.

\bibitem[{Braatz et~al.(1994)Braatz, Young, Doyle, and
  Morari}]{braatz1994computational}
Braatz, R.P., Young, P.M., Doyle, J.C., and Morari, M. (1994).
\newblock Computational complexity of/spl mu/calculation.
\newblock \emph{IEEE Trans. on Automatic Control}, 39(5), 1000--1002.

\bibitem[{Doyle(1982)}]{doyle1982analysis}
Doyle, J. (1982).
\newblock Analysis of feedback systems with structured uncertainties.
\newblock 129(6), 242--250.

\bibitem[{Dubanchet(2016)}]{DV16}
Dubanchet, V. (2016).
\newblock \emph{{Modeling and control of a flexible Space robot to capture a
  tumbling debris}}.
\newblock Ph.D. thesis, ISAE-SUPAERO, Polytechnique Montr\'eal.

\bibitem[{Thai et~al.(2019)Thai, Roos, and Biannic}]{ThRoBi19_acc}
Thai, S., Roos, C., and Biannic, J.M. (2019).
\newblock Probabilistic $\mu$-analysis for stability and $\mathcal{H}_\infty$
  performance verification.
\newblock In \emph{Proceedings of the ACC}, 3099--3104.

\bibitem[{Zhou and Doyle(1998)}]{zhou1998essentials}
Zhou, K. and Doyle, J.C. (1998).
\newblock \emph{Essentials of robust control}.
\newblock Prentice hall Upper Saddle River.

\end{thebibliography}
\end{document}